# The SPORT-C Intervention: An Integration of Sports, Case-Based Pedagogy and Systems Thinking Learning

**Jeffrey Basoah[1]· Dr. William Scherer[2] · Dr. Karis Boyd-Sinkler[3] · Dr. Reid Bailey[4]**

**Abstract** The STEM field is unrepresentative of the population it serves. Due to a lack of cultural relevance in STEM courses, there is a dissociation between the lived experience of students from underrepresented racial groups (URG) and STEM course material. The SPORT-C intervention is a framework that combines sports, systems thinking learning, and a case-based pedagogy into an activity that can be used in any STEM course. A pilot study was conducted to determine the viability of the SPORT-C intervention in a classroom setting and determine if it was worth further investigating and if any impact differed by racial identity. The findings from this study implicate that the SPORT-C intervention has an impact on the motivation levels of students to participate in STEM courses.

## 1 Introduction

The STEM workforce needs racial diversity. Those who identify as Black or African American, Hispanic or Latino, American Indian or Alaska Native, and Native Hawaiian or Other Pacific Islander are underrepresented in the workforce, collectively making up 15.1% of the STEM workforce despite making up 31.9% of the United States (U.S.) population (Burke & Okrent, 2021; US Census Bureau, 2020); these races and ethnicities are further referred to as underrepresented racial groups (URGs). In contrast, while Whites make up 66.2% of the STEM workforce, they only make up 60.1% of the U.S. population; Asians make up 16.3% of the workforce but just 5.8% of the U.S. population (Burke & Okrent, 2021; US Census Bureau, 2020); these races and ethnicities are further referred to as overrepresented racial groups (ORGs). Lack of representation in the STEM workforce can lead to innovations that are ineffective at addressing problems that adhere to the needs of all people (Quiroz-Rojas & Teruel, 2021). STEM education serves as a pipeline for diversifying the STEM workforce (Palmer et al., 2017). Previous research has established the importance of early interest in STEM as a critical factor in remaining persistent in the field (Tai et al., 2006). The present pipeline fails to capture youth for whom STEM remains remote and outside their current experience (Drazan, 2020). This study proposes and tests a novel approach to addressing lack of representation in STEM education through the innovative integration of sports and case-based systems thinking learning.

[1] Department of Engineering Systems and Environment, University of Virginia, 4201 Sonia Ct, Alexandria, VA 22309, United States of America, Email: jkb2jf@virginia.edu

[2] Department of Engineering Systems and Environment, University of Virginia, 151 Engineers Way, Charlottesville, VA 22903, United States of America, Email: wts@virginia.edu

[3] Pratt School of Engineering, Duke University, 5010 Garrett Road, Apt 204, Durham, NC 27707, United States of America, Email: karisboyd@gmail.com

[4] Department of Engineering Systems and Environment, University of Virginia, 151 Engineers Way, Charlottesville, VA 22903, United States of America, Email: rrb5b@virginia.edu

## 2 Background

**STEM Education and the STEM Field**  Studies have shown how STEM career aspirations are linked to the pursuit of STEM education (Maltese & Tai, 2011). The majority of students who focus on STEM do so in high school, and their decision is based on a growing interest in math and science (Maltese & Tai, 2011). Efforts to increase the number of URG students by influencing their career aspirations during adolescence could help shape the sector's future (Kohen & Nitzan, 2022). Given the country's rapidly shifting ethnic demographics, increased access and opportunities for URG students to seek, persist in, and succeed in STEM education will be critical in establishing a diverse STEM workforce in the United States (Palmer et al., 2017).

**Educational Context**  According to the existing literature, one effective way to increase students' STEM career interest is to communicate the personal and social values of STEM and the connections between STEM and the real world (Sheldrake et al., 2019). A disconnect between students and their learning context can also lead to a drop in motivation and academic performance, subsequently affecting aspirations within that field of context (Sheldrake et al., 2017). Culturally relevant pedagogy strongly suggests that a student's connection to the material taught in their courses is essential for academic success (Ladson-Billings, 2014). Other prevailing theories such as cultural mismatch (Stephens et al., 2012), identity-based motivation (Oyserman & Destin, 2010), and goal congruity (Diekman et al., 2010) suggest that alignment between a student's background, identity, values, or goals and the educational context serve as successful motivational resources. Developing students' attitudes, and hence their aspirations, by highlighting science's applications and relevance to everyday life may be beneficial in drawing interest to the field.

**Sports**  Sports has been a demonstrated unique approach to engaging students in STEM education as recent studies have investigated the effects of linking sports concepts with STEM learning (Ali et al., 2021). In 2019, data indicated that over half of high school students participate in school sports, 73% of students ages 6 – 12 and 69.1% ages 13 - 17 (Aspen Institute, 2020a, 2020b; Wretman, 2017). With more than one out of every two students participating in sports activity, it is safe to say most students have interests in this domain. 34.8 % of Black and 34.1 % of Hispanic children aged 6 to 12 played a sport, while 42.4 % of Black and 40.3 % of Hispanic children aged 13 to 17 played a sport[5] (Aspen Institute, 2020a, 2020b). These figures represent the lowest estimate of the number of students interested in sports. Students who are interested in sports but cannot participate due to a variety of reasons such as a lack of availability at school are not included. With 7.1 % of Black and 7.7 % Hispanic workers in the STEM workforce, there is an opportunity to increase participation by these URGs using sports as a bridge (Burke & Okrent, 2021).

**Case Based Pedagogy**    A case is a description of an actual situation, commonly involving a decision, a challenge, an opportunity, a problem, or an issue faced by a person or persons in an organization, requiring the reader to step figuratively into the position of a particular decision-maker (Herreid, 2007). Case-based education can make the curriculum more relevant and motivational for students by challenging them to apply what they've learned to real-life situations. Multiples studies have demonstrated the effectiveness of a case-based instructional pedagogy. Case studies have been found to increase students' critical thinking and problem-solving skills (Akili, 2012), self-efficacy in subject material (Holley, 2017), and subject comprehension (Munakata-Marr et al., 2009). Findings from studies centered around the effects of a case-based pedagogy suggest that improving kids' enjoyment, interest, and perceptions of their STEM skill, as well as their judgments of its worth in a future STEM career, may lead to an increase in the number of students studying STEM in school.

**Systems Thinking Learning**    Systems thinking is comprehending complex interconnected inputs and outputs that work toward a common goal while analyzing problems (Lavi & Dori, 2019; York et al., 2019). Systems thinking utilizes a holistic approach, which encompasses tackling the problem as a whole versus individual parts (Camelia & Ferris, 2016; Hossain et al., 2020). Systems thinking has received considerable attention in recent years, with how to teach systems thinking, including how to teach it to K-12 students (Lavi & Dori, 2019). Educators have already seen the value of incorporating systems thinking into their classrooms and how it prepares students to make informed, ethical decisions about relevant issues (Delaney et al., 2021). Educators who have used systems thinking in their classroom have reported that students are active participants in their learning, learn content more deeply and conceptually, ask better questions, and make more connections between concepts both within and between disciplines (York et al., 2019).

**The SPORT-C Intervention** The SPORT-C intervention is a framework in which sports, a case-based pedagogy, and systems thinking learning are blended into an activity potentially applicable in all STEM courses (see Figure 1). With the SPORT-C intervention, sports are seamlessly incorporated into the classroom curriculum in the form of a sports case. Students will be given a real-world scenario centered around sports that will incorporate the learning objectives for whichever unit they are in their curriculum. The students will need to utilize their current unit's learnings to progress through the activity while simultaneously use the systems thinking problem-solving method to complete the activity.

---

[5] Native American data was not reported, while Asian and Pacific Islander data were reported together.

**Fig. 1** The Sport-C intervention diagram. Displays how each component works together to foster STEM learning

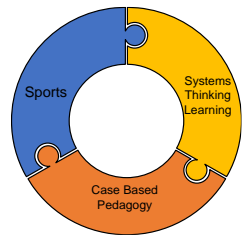

**Purpose of Study** To test the effectiveness of the SPORT-C intervention, a pilot study was conducted to determine its viability in a classroom environment and to see if it was worth exploring further. The research questions that guided this study are: *RQ1: How does the SPORT-C intervention impact a student's motivation to participate in their STEM course? RQ2: Does the impact of the SPORT-C intervention vary by racial identity?*

The motivation was measured by analyzing a participant's self-reported academic engagement, self-efficacy, expectancy, value, and cost of STEM-related courses before and after students were exposed to the SPORT-C intervention (Lazowski & Hulleman, 2016). The following is how each measured factor is defined in the context of this study:

| | |
|---|---|
| Academic engagement | Degree of attention that an individual shows when they are learning (Great Schools Partnership, 2013) |
| Self-efficacy | Individuals confidence in their ability to successfully complete tasks (Luo et al., 2020) |
| Expectancy | Extent to which a student thinks he or she can be successful in a task (Kosovich et al., 2015) |
| Value | Extent to which a student thinks a task is worth completing (Kosovich et al., 2015) |
| Cost | Negative aspects of participating in an activity, or the loss of other valuable activities (Kosovich et al., 2015) |

In addition, study participants took part in a focus group to gain a qualitative understanding of their learning experience while undergoing the SPORT-C intervention. A brief interview with the classroom instructor (CI) was conducted to obtain a different perspective on the impact of the SPORT-C intervention on student motivation levels.

**Positionality Statement** To begin this conversation, it is helpful to understand the author's positionality and the lens they are viewing the data. The sole researcher is an African-American U.S.-born scholar with immigrant parents from Ghana. The researcher is a 2nd year Master's student studying systems engineering at a university with a predominately white student body and faculty. They earned their bachelor's degree in mechanical engineering from a different university with a predominately white student body and faculty. It is likely that the racial background of the author may have influenced the interpretations and implications of the study. To mitigate the influence of the author's background and provide reflexivity to the study, the author had an external audit conducted by a third party on the research to evaluate the accuracy and whether or not the findings, interpretations, and conclusions are supported by the data (Lincoln & Guba, 1985).

# 3 Methods

**Data Collection** The participants in this study were students at a public high school in the southeast region of the United States. The high school hosts grades 9 -12 and has a 12:1 student-teacher ratio with under 1200 students. The school has a demographic breakdown of 50.7% ORG enrollment (44.5% White and 6.2% Asian) and 49.3% URG (28.7% Black, 13.6% Hispanic, 7.0% two or more races, Native American or American Indian and Hawaiian and Pacific Islander not reported) with 51% identifying as male and 49% identify as female (Virgina.gov, 2022). Participants were all enrolled in a multi-grade level math class, "Algebra, Function, and Data Analysis" (AFDA). Participants for the study were selected based on their class period. The control group was pulled from 11 students who attended Period 3 AFDA, the intervention group was pulled from 21 students who attended Period 6 AFDA. The study was conducted in March 2022, and Covid-19 cases were still prevalent in the area; however, the class was still held in person. As noted by the CI, class attendance was inconsistent due to the school policy for students to remain at home if experiencing any COVID-like symptoms. The study began with 15 participants, 11 from the experimental group and four from the control. Only 10 participants completed the study due to unavailability: three from the control group and seven from the experimental group.

**Procedure** Before receiving any study material, the participants all engaged in the same classroom lesson administered by the same CI on using probability to calculate expected values during their respective class periods. At the end of their respective lessons, both intervention and control groups received their respective pre-surveys. The following class period, both groups received their activity. After completing the activity, the participants from both groups were asked to complete a post-activity survey. The post- activity survey was the same as the control group pre-survey. Participants from both groups completed the post-survey the class period following their activity day due to running out of class time. Focus group participants were gathered and interviewed two class periods, or three days, after the intervention activity

**Intervention and Control Group Activity** Participants were exposed to the SPORT-C intervention activity in the intervention group. The researcher designed and prepared the SPORT-C intervention in collaboration with the classroom instructor to ensure that the lessons learned were comparable to those learned by the control group. The intervention was developed so that no special or external training would be required of the CI. The SPORT-C intervention activity involved the participants using probability and expected values to solve a scenario that could present itself in the basketball world. Basketball was specifically chosen as the activity due to the familiarity with the sport by the CI, and the instructor felt that a specific basketball scenario would coincide well with their current lesson plans. The control group participated in their scheduled non-sports related classroom activity. The control activity had no central theme. Participants were asked to calculate the expected value for investment portfolios, various games involving probability, and ticket sales. The CI allowed participants in each group to work in groups of three.

**Survey Development** A 40-item survey was used to gather quantitative data on the participant's motivation levels before and after the respective activities. A series of validated survey scales were utilized to construct the entirety of the survey and then administered to participants. For instrument reliability, chosen scales had a minimum Cronbach αor McDonald's ω of .7 or higher (Streiner, 2003). The survey asked five questions about students' academic engagement (Leibowitz et al., 2020), 12 questions about self-efficacy (Luo et al., 2020), three questions about expectancy (Kosovich et al., 2015), three questions about values (Kosovich et al., 2015), four questions about the cost of STEM courses (Kosovich et al., 2015), and 13 demographic questions.

**Focus Group and Interviews** A focus group was conducted to encourage participants to share their individual experiences about the intervention activity holistically. Only participants from the experimental group were recruited to participate in a follow-up focus group. Any participation from control group participants would not have been advantageous for analysis as they were not exposed to the intervention activity. The focus group consisted of three participants. Initially, four had volunteered to participate; however, only three were present on the session day. A semi-structured interview was conducted via Zoom. Participants were asked questions ranging from what they liked and disliked about the activity, what they would change if given the opportunity, and the sports topic's impact on their learning experience. The CI interview was conducted to provide qualitative data on the participants' engagement from the instructor's perspective. This provided the study with insight into the perceived effectiveness from the instructor's lens and data on the progression of the activity during its dissemination.

**Data Analysis Procedure** To perform any quantitative analysis on the survey responses, the pre and post responses were matched using student identifiers given to the participants at the start of the study. Any responses with more than two missing items from a subscale had their subscale removed from the analysis. First, pre and post-survey subscale scores were calculated by averaging Likert responses for each factor. Then, a repeated measures multifactor ANOVA test was conducted to measure a relationship between the intervention activity and participant motivation levels. The dependent variable for the analysis was the mean scores for participant response to subscales. The average subscale score for participants pre- and post-survey (time) and their condition (intervention or control group) were used as the independent variables. To determine if each participant pool was appropriate for comparison, a baseline equivalence test was performed. To determine baseline, a multifactor ANOVA test was performed on each subscale to determine the between subject effects. The independent variables were the control and intervention groups, the dependent variable was the mean of a subscale before and after the activities.

A grounded theory method for the coding process was invoked to perform qualitative analysis on the focus group and CI interview (Birks & Mills, 2011). The coding process and analysis were completed in four cycles for each session, and an analytical memo was developed for each as well.

## 4 Results

**Quantitative Results** A summary of the difference in mean values from pre- to post-survey responses in both the control and intervention groups by racial group is shown in Table 1. Blue shaded boxes indicate a positive difference in mean values relative to scales, and yellow shaded boxes indicate a negative change. White boxes indicate no change. Due to incomplete survey responses, some analyses for EXP, VAL, and COS subscales were unable to be performed by the researcher. Table 1 also displays the p-values for the baseline equivalence test. The self-efficacy and value subscales were the only subscales to output a significant difference between the experiment and control group.

**Table 1.** Mean value differences for subscales and p values from baseline equivalence test results

| | Control | | | | Intervention | | | | Baseline equivalence |
|---|---|---|---|---|---|---|---|---|---|
| | ORG | N | URG | N | ORG | N | URG | N | p value |
| Academic Engagement | -0.60 | 1 | 0.40 | 2 | 1.00 | 1 | 0.54 | 6 | .392 |
| Self-Efficacy | 0.25 | 1 | -0.08 | 2 | -0.15 | 1 | 0.07 | 6 | .027 |
| Expectancy | 0.67 | 1 | 0.50 | 2 | - | 0 | 0.20 | 5 | .274 |
| Value | -0.67 | 1 | 0.33 | 2 | - | 0 | -0.20 | 5 | .007 |

| Cost | 0.00 | 1 | -0.50 | 2 | - | 0 | -0.85 | 5 | .161 |
|------|------|---|-------|---|---|---|-------|---|------|

**Qualitative Results** The following sections are organized by the broader themes/categories developed during the coding phase. Over the six final codes and subcodes developed, three themes emerged: learning experience, structure, and relevancy.

*Theme 1 – Learning experience* The main objective of the focus group was to provide the participants with another opportunity to share their experiences while participating in the SPORT-C intervention activity and address RQ1. Within this theme, three subthemes are included that pertain specifically to the participant's learning experience: activity engagement, activity difficulty, and activity comprehension.

Both the participants and the CI expressed levels of engagement higher than the norm. Participants stated several times that they felt more engaged while participating in the activity. For example, participant 2 said, "I was actually paying attention in class for once which is rare for me" and they that they "didn't zone out completely". Their responses are echoed by the CI as they mentioned the students "definitely seemed interested" and that "everybody was pretty engaged, even the kids that are usually little harder to get involved they seemed like they all put their head down and got their work done." The CI went as far as to mention that two students in particular that require a bit of their attention to get engaged with the classwork "seem[ed] to do pretty well the whole time". There was no mention of disinterest from the students or the CI.

There was a mixture of responses regarding the activity difficulty for the participants. Participants all agreed that the format of the SPORT-C intervention made it easier for them to complete the assignment. Multiple participants stated that the assignments were easy but posed a challenge for them. Participant 1 voiced the "roller coaster ride" of difficulty stating, "For this worksheet I felt as if it was hard to get, of course, because [it was] something new, but as time went on it got easier, but then also got harder again, but then it got easier, you know it wasn't just consistent with how hard it was getting". This "roller coaster" ride was coupled with a sense of accomplishment, later discussed. It is also key to note that Participant 1, in particular, expressed that their only disgruntle with the SPORT-C intervention was its difficulty level in the beginning, stating, "I didn't really have much complaints aside from the starting factor of how hard it was at first until it just got easier."

All three participants echoed that they could comprehend what they were working on. Participant 1 notably stated that "[I] felt like I was actually like understanding some of the things that I was actually working on" while comparing the SPORT-C intervention to the classroom's routine activities. Participant 3 stated some confusion during the expected value lesson before the intervention activity. They later gained understanding after the intervention activity was implemented stating, "For me [it] was good [as] basketball is my favorite sport and to use it [it] make more sense for me to do because before we started it was very hard [but] like when he explained to us I understand a lot of what he was doing". The participants expressed more confidence in their ability to do and complete the intervention activity. Participant 2 reflected upon his experience while completing the intervention activity, "You know I learned a few things by myself, even though I got stuck really on most a lot of places, but [as] it just gradually went through my head I collected myself, and you know I just push[ed] through without any help at all, which you know I actually you know I really loved about it to be honest."

*Theme 2 – Structure* A prominent topic of discussion was the overall structure of the SPORT-C intervention. The structure does not address RQ1 however provides the researcher insights into the reception of the intervention for future iterations. Participants frequently mentioned their feelings on the structure compared to their normal classroom flow. All students preferred the SPORT-C intervention over the normal classroom flow as "[it] just felt more put together," stated participant 1 several times. Participants 2 and 3 took to the structure of SPORT-C intervention as they appreciated the guidance provided by the CI throughout its entirety. After expressing earlier in the focus group that they were "more of an independent worker kind of person" who tended to "rush ahead" during assignments, Participant 1 stated that structure allowed them to work independently while remaining attentive to the CI's guidance, something they would not have done previously.

*Theme 3 – Relevancy* This theme relates to RQ1 as relevancy is correlated with value and can influence student motivation if present. All participants said they were familiar with basketball as a spectator of the sport or as a recreational activity. Participants expressed being able to make connections "quicker" as the topic was something they were all familiar with. For example, regarding basketball, participant 1 stated, "Like it was something real. I understand [the assignment] quicker because it's something you know that's like in our world and tangible and like is an actual something that a lot of people watch and participate in, so I think that just made the connection easier in my brain." Participant 2 echoed the same sentiment as being able to make "sense faster" of the assignment due to both the topic and the structure of the intervention.

# 5 Discussion

The quantitative data was collected to help answer both RQ1 and RQ2, and the qualitative data was collected to help answer both RQ1 and RQ2, but due to the small sample size, the data has no statistical significance. While some of these results suggest that the intervention was beneficial, they must be viewed as preliminary and not generalizable. To answer RQ1, the SPORT-C intervention impacts student motivation to participate in STEM courses, based on focus group responses. Since the ORG intervention group did not respond to the EXP, VAL, or COS subscales, the answer to RQ2 is based solely on a non-generalizable interpretation of the AE and SE subscale scores. As the study did not have an adequate sample size and the focus group participants were de-identified, the answer to RQ2 is inconclusive.

It's interesting to see how URGs exhibited a downward trend in the intervention group's VAL subscale scores from pre- to post-survey, but a positive trend in the control group. Because both groups fail the baseline equivalence test for that subscale, there is no way to compare their VAL subscale score changes. The increase in AE subscale scores corresponds to the CI and focus group responses. There was general agreement that there was a higher level of engagement while participating in the activity. Participants noticed changes in their usual behavior and the CI in their students due to the SPORT-C intervention. The intervention did not require any additional motivation from normally disengaged participants or required special attention. Some participants who would normally require the CI's assistance to jumpstart classwork did not require it during the intervention activity, surprising the CI.

It's worth noting that URGs in the intervention group saw a reduction in cost. This indicates that the participants did not view the intervention activity as a waste of time. This finding coincides with the focus group's assessment of the activity's difficulty and participation. The participants stated that completing the task was difficult at first. It's natural for students to avoid work that becomes too difficult. However, as one participant pointed out, the intervention's difficulty only encouraged them to try harder. The participant's perception of the activity's difficulty prompted them to work harder on the assignment and overcome the challenge. The participant finished the activity with pride in their ability to complete it. As a result of the intervention, the students' self-efficacy and expectancy in STEM activities increased. The intervention's difficulty level in the activity was not studied, but it is something to keep in mind for future iterations. The students sacrificed more time to complete the activity as they put more effort into it. They were rewarded with a sense of accomplishment despite putting in more time than usual to complete the activity. This tradeoff is reflected in the participant's cost response.

## 6 Conclusion

Through the innovative integration of sports and case-based systems thinking learning, this study proposed and tested an approach to addressing social inequity in STEM education. The findings from this study imply that the SPORT-C intervention impacts the motivation levels of students to participate in STEM courses, but no generalizable quantitative or qualitative data was produced. Participants reported higher levels of engagement, a sense of accomplishment from completing perceived complex tasks, a quicker grasp of the learning topic due to relevance, and a preference for the intervention's activity structure. The findings support the researcher's decision to continue investigating the interventions' impact on student's motivation to pursue STEM courses as means to diversify a field that isn't representative of the population it serves.